\documentclass[a4paper,12pt]{article}
\pdfoutput=1 
\usepackage[english]{babel}
\usepackage{amsmath}
\usepackage{amssymb}
\usepackage{bbm}
\usepackage{color}
\usepackage{cite}
\usepackage{xcolor,hyperref}
\usepackage{graphicx}
\usepackage[titletoc,toc,title]{appendix}
\usepackage[font=footnotesize,labelfont=bf]{caption}
\Roman{section}
\newcommand{\naw}[1]{\left(#1\right)}

\newcommand{\modu}[1]{\left|#1\right|}
\newcommand{\poisson}[1]{\left\{#1\right\}}

\title{Kepler problem and chiral effective dynamics}
\author{Katarzyna Bolonek-Laso\'n$^1$\footnote{katarzyna.bolonek@uni.lodz.pl} \and Joanna Gonera$^2$\footnote{joanna.gonera@uni.lodz.pl} \and Piotr Kosi\'nski$^2$\footnote{piotr.kosinski@uni.lodz.pl} }
\date{%
	\small{$^1$Department of Statistical Methods, Faculty of Economics and Sociology\\ University of Lodz, 41/43 Rewolucji 1905 St., 90-214 Lodz,  Poland}\\%
	$^2$Department of Computer Science, Faculty of Physics and Applied Informatics \\University of Lodz, 149/153 Pomorska St., 90-236 Lodz, Poland }

\begin{document}
	\maketitle
	
\begin{abstract}
It is shown that by an appropriate canonical transformation Kepler dynamics can be put in the form which allows to exhibit the structure of the symmetry transformations related to the superintegrability. They appear to fit nicely into general scheme of nonlinear realizations. In new coordinates the Kepler dynamics results from dimensional reduction of that describing low energy mesons with spontaneously broken chiral symmetry.			
\end{abstract}

\section{Introduction}
The Kepler \cite{Guillemin}, \cite{Cordani} problem may be viewed as one of the pillars of Hamiltonian dynamics. It provides good approximation for dynamics of solar system and other planetary systems. From theoretical point of view it is distinguished by the high degree of symmetry: it is not only integrable in the Arnold-Liouville sense \cite{Arnold} but also maximally superintegrable - it admits maximal number (five) of functionally independent globally defined integrals of motion. 

Rich symmetry structure makes the Kepler problem ideal laboratory for studying the power and effectiveness of group theoretical methods (for extensive discussion and bibliography see, for example, Refs. \cite{Cordani} and \cite{Bohm}).

The intriguing properties of Kepler dynamics were discussed in numerous papers starting from those of Fock \cite{Fock} and Bargmann \cite{Bargmann} who revealed the structure of symmetries underlying superintegrability of Kepler dynamics.

In the present paper we discuss the global structure of these symmetry transformations. We show that by a canonical transformation one can define new variables in terms of which the symmetry generated by conserved quantities takes the standard form of nonlinear realization of $SO(4)$ group linearising on rotation subgroup. The resulting Lagrangian may be viewed as dimensional reduction of the effective Lagrangian describing pions as Goldstone bosons.

Let us conclude the introductory section by recalling in some detail the original Kepler problem and its symmetries. One considers a point particle of mass $m$ moving in attractive central potential inverse proportional to the distance from the origin. The relevant Hamiltonian reads 
\begin{equation}
	\mathcal{H}=\frac{\vec{p}^2}{2m}-\frac{k}{r},\qquad r\equiv\modu{\vec{x}}\label{d1}
\end{equation}	
with $k>0$ being the coupling constant. The Hamiltonian \eqref{d1} does not depend explicitly on time which implies time translation symmetry and, via Noether theorem, energy conservation. Moreover, due to the rotational invariance the (orbital) angular momentum is conserved as well
\begin{equation}
	\vec{L}\equiv\vec{x}\times\vec{p}, \qquad \dot{\vec{L}}=0.\label{d2}
\end{equation} 
These symmetries and conservation laws are shared by all central conservative potentials. However, due to the particular form of $r$-dependence, there are additional integrals of motion in the Kepler problem. Namely, one can define the so-called Runge-Lenz vector
\begin{equation}
	\vec{A}\equiv\vec{p}\times\vec{L}-mk\frac{\vec{x}}{r}\label{d3}
\end{equation}
which is also conserved:
\begin{equation}
	\dot{\vec{A}}=0.\label{d4}
\end{equation}
$\vec{A}$ obeys the following relations
\begin{equation}
	\vec{A}\cdot\vec{L}=0\label{d5}
\end{equation}
\begin{equation}
	\vec{A}^2=m^2k^2+2mE\vec{L}^2,\qquad E=\mathcal{H}.\label{d6}
\end{equation}

Summarizing, Kepler dynamics exhibits seven integrals of motion, $\mathcal{H}$, $\vec{L}$ and $\vec{A}$, obeying two constraints \eqref{d5}, \eqref{d6}. As a result, we obtain five functionally independent integrals of motion yielding Kepler dynamics superintegrable.

The integrals $\mathcal{H}$, $\vec{L}$, $\vec{A}$ obey nice Poisson commutation rules
\begin{equation}
	\poisson{L_i,L_j}=\varepsilon_{ijk}L_k \label{d7}
\end{equation}
\begin{equation}
	\poisson{L_i,A_j}=\varepsilon_{ijk}A_k \label{d8}
\end{equation}
\begin{equation}
	\poisson{A_i,A_j}=-2m\mathcal{H}\varepsilon_{ijk}L_k. \label{d9}
\end{equation}
On the energy hypersurfaces \eqref{d7}$\div$\eqref{d9} define the Lie algebra structures: $SO(4)$, $e(3)$, $SO(3,1)$ for $\mathcal{H}<0$, $\mathcal{H}=0$ and $\mathcal{H}>0$, respectively.

\section{Kepler dynamics from $SO(4)$ coadjoint orbit}

The natural tool for describing the phase spaces of Hamiltonian systems exhibiting symmetry is provided by the notion of coadjoint orbits of symmetry group \cite{Arnold}, \cite{Kostant,Kirillov1,Kirillov2,Woodhouse,Soriau1,Marsden}. 
It is, therefore, not surprising that, due to its rich symmetry structure, the Kepler Hamiltonian system can be described in terms of coadjoint orbit of certain group ($SO(4,2)$ as will be described below). Before entering the particular case of Kepler system we briefly sketch, for reader's convenience, the main points of the coadjoint orbit method (cf., for example, Ref.~\cite{Arnold}). From the physical point of view it provides the solution to the following problem. Given a Lie group $G$ we want to construct (and classify) the phase spaces (i.e. even-dimensional manifolds equipped with non-degenerate Poisson brackets) on which $G$ acts as the group of canonical transformations (i.e. leaving Poisson brackets invariant). It appears that, under the additional assumption of transitive action of $G$, the relevant phase spaces are (up to some topological subtleties) the coadjoint orbits of $G$. Let $\poisson{X_i}_{i=1}^n$ be the set of generators of the group $G$. Any element of the Lie algebra $\mathcal{G}$ of $G$ can be written as $X=\sum\limits_{i=1}^n\zeta^iX_i$, $\zeta^i\in\mathbbm{R}$. 

Let $\{\tilde{X}^i\}_{i=1}^n$ be the dual basis in the dual space $\mathcal{G}': \,\langle\tilde{X}^i,X_j\rangle={\delta^i}_j$. Then any $\tilde{X}\in \mathcal{G}'$ can be written as $\tilde{X}=\sum\limits_{i=1}^n\zeta_i\tilde{X}^i$, $\zeta_i\in\mathbbm{R}$. In $\mathcal{G}'$ one can define the natural (but degenerate) Poisson structure
\begin{equation}
\poisson{\zeta_i,\zeta_j}={c_{ij}}^k\zeta_k\label{b1}
\end{equation}  
with ${c_{ij}}^k$ being the structure constants of $G$ ($[X_i,X_j]=i{c_{ij}}^kX_k$). Let $Ad_g(X)$ be the adjoint action of $G$ on $\mathcal{G}$. The coadjoint action of $G$ on $\mathcal{G}'$ is defined by
\begin{equation}
\langle Ad_g^*(\tilde{X}),Y\rangle=\langle\tilde{X},Ad_g(Y)\rangle,\quad \tilde{X}\in\mathcal{G}',\quad Y\in\mathcal{G}.\label{b2}
\end{equation}
Now, the important points are: (\textit{i}) the Poisson brackets \eqref{b1} are invariant under the coadjoint action of $G$; (\textit{ii}) the Poisson brackets become nondegenerate on coadjoint orbits of $G$ in $\mathcal{G}'$; (\textit{iii}) essentialy, all phase spaces on which $G$ acts transitively as a group of canonical transformations are coadjoint orbits.

Summarizing, the coadjoint orbits method allows us to construct all Hamiltonian systems with a given transitively acting, group of canonical transformations. For example, in this way one can classify all elementary Hamiltonian systems obeying relativity principle, both in Galilei and Einstein form. 

In order to find the group-theoretical background of Kepler problem one has to regularize it. In fact, the Kepler dynamics is not regular because the vector field generated by the Hamiltonian \eqref{d1} is not complete: for the orbits corresponding to the vanishing angular momentum the particle gets to the attractive center in finite time with infinite velocity. Once the regularization is performed one finds that the full dynamical group of Kepler problem is $SO(4,2)$ \cite{Marsden,Moser,Soriau2,Onofri,Onofri1,Kummer1,Cordani1,Meer}.
Therefore, one should describe Kepler dynamics in terms of coadjoint orbits.

 The generic (co)adjoint orbit of $SO(4,2)$ is twelve-dimensional (because $SO(4,2)$ is fifteen-dimensional group of rank three) 
 while we need six-dimensional phase space which implies that one has to consider nongeneric (singular orbit). We start with dual space to the Lie algebra of $SO(4,2)$. Denote its coordinate functions by $\zeta_{ab}=-\zeta_{ab}$, $a,b=0,1,2,3,5,6$ (we adopt here the Todorov convention \cite{Todorov} omitting index 4). The basic Poisson brackets read 
\begin{equation}
	\poisson{\zeta_{ab},\zeta_{bc}}=g_{ad}\zeta_{bc}+g_{bc}\zeta_{ad}-g_{ac}\zeta_{bd}-g_{bd}\zeta_{ac}\label{a1}
\end{equation}
where $g_{ab}=diag(+----+)$.

The singular six-dimensional orbit relevant in the present context can be defined by $SO(4,2)$ - covariant equation \cite{Kosinski}
\begin{equation}
	{\zeta_a}^c\zeta_{cb}=0.\label{a2}
\end{equation}
Assuming Greek letters run from 1 to 5 (except 4) and putting
\begin{equation}
	\zeta_{0\mu}\equiv \omega_\mu,\quad \zeta_{6\mu}\equiv z_\mu \label{a3}
\end{equation}
one finds from \eqref{a2}
\begin{equation}
	\omega_{\mu}\omega_\mu=z_\mu z_\mu\label{a4}
\end{equation}
\begin{equation}
	\omega_\mu z_\mu =0\label{a5}
\end{equation}
\begin{equation}
	\zeta_{06}=\pm\sqrt{\omega_\mu\omega_\mu}=\pm\sqrt{z_\mu z_\mu}\label{a6}
\end{equation}
\begin{equation}
	\zeta_{\mu\nu}=\frac{1}{\zeta_{06}}\naw{\omega_\mu z_\nu-\omega_\nu z\mu}.\label{a7}
\end{equation}
Eqs.~\eqref{a3}$\div$\eqref{a7} provide the complete description of six-dimensional orbit of $SO(4,2)$; in what follows we will choose "+" sign in eq.~\eqref{a6}.

It is interesting to view this orbit as a group coset $SO(4,2)/G_{\underline{\zeta}}$ with $G_{\underline{\zeta}}$ being the stability subgroup of some "canonical" point on the orbit under consideration. It is not difficult to find such a point $\underline{\zeta}$ and determine $G_{\underline{\zeta}}$ on the infinitesimal level. However, the description of the global structure of $G_{\underline{\zeta}}$ is more involved. Fortunately, a convenient choice of $\underline{\zeta}$ and the global structure of $G_{\underline{\zeta}}$ has been described in some detail by Onofri and Pauri \cite{Onofri2}. $G_{\underline{\zeta}}$ appears to be the semidirect product of two groups, the first being the direct product of $SO(2)$ and $SO(2,1)$ while the second - five-dimensional Lie group of Euclidean topology. We refer to \cite{Onofri2}, eq.~(66)
and further, for more details.

 The initial Poisson structure \eqref{a1} becomes now nondegenerate and takes the form
\begin{equation}
	\poisson{\omega_\mu,\omega_\nu}=-\frac{1}{\zeta_{06}}\naw{\omega_\mu z_\nu-\omega_\nu z_\mu}\label{a8}\end{equation}
\begin{equation}
	\poisson{z_\mu,z_\nu}=-\frac{1}{\zeta_{06}}\naw{\omega_\mu z_\nu-\omega_\nu z_\mu}\label{a9}
\end{equation}
\begin{equation}
	\poisson{\omega_\mu,z_\nu}=\zeta_{06}\delta_{\mu\nu}.\label{a10}
\end{equation}
It can be shown \cite{Gyorgyi}, \cite{Onofri} that there exist globally defined Darboux canonical variables and the Hamiltonian expressible in terms of $\zeta_{06}$ coordinate function which describe the Kepler problem; the actual form of the transformation from $\omega_{\mu}$, $z_\mu$ to canonical ones (the so-called Bacry-Gyorgyi transformation) is, however, quite complicated (cf.~eqs.~\eqref{a8} in Ref.~\cite{Onofri}). Nevertheless, one can conclude that $SO(4,2)$ is the dynamical group of Kepler problem.

$SO(4,2)$ is also the conformal symmetry group of Minkowski spacetime. This strongly suggests that there exists a connection between Kepler dynamics and conformal geometry of Minkowski spacetime \cite{Keane}, \cite{Cariglia}. The six-dimensional coadjoint orbit of $SO(4,2)$, defined by eq.~\eqref{a2}, describes relativistic particle with vanishing mass and helicity \cite{Kosinski}. The Poincare symmetry of such a particle can be easily extended to the conformal one and the generators of $SO(4,2)$ are expressible in terms of $\omega_\mu$ and $z_\mu$. Moreover, for vanishing helicity (and only in this case \cite{Skargerstam}) one can find the global Darboux variables related in the natural way to the description of point particles \cite{Kosinski}, \cite{Skargerstam}. They read \cite{Kosinski}:
\begin{equation}
	x_i\equiv -\frac{\omega_i}{\omega_5+\zeta_{06}}=-\frac{\omega_i}{\omega_5+\sqrt{\omega_\mu\omega_\mu}}\qquad i=1,2,3\label{a11}
\end{equation}
\begin{equation}
	p_i\equiv \frac{z_5}{\zeta_{06}}\omega_i-\naw{\frac{\omega_5}{\zeta_{06}}+1}z_i=\frac{z_5}{\sqrt{\omega_\mu\omega_\mu}}\omega_i-\naw{\frac{\omega_5}{\sqrt{\omega_\mu\omega_\mu}}+1}z_i.\label{a12}
\end{equation}
On the orbit under consideration all coordinate functions $\zeta_{ab}$ can be expressed in terms of $\vec{x}$ and $\vec{p}$:
\begin{equation}
	\zeta_{ij}=x_ip_j-x_jp_i\label{a13} 
\end{equation}
\begin{equation}
	\zeta_{0i}=-\modu{\vec{p}\,}x_i\label{a14} 
\end{equation}
\begin{equation}
	\zeta_{56}=\vec{x}\cdot\vec{p}\label{a15}
\end{equation}
\begin{equation}
	\zeta_{05}=\frac{\modu{\vec{p}\,}}{2}\naw{1-\vec{x}\,^2}\label{a16} 
\end{equation}
\begin{equation}
	\zeta_{06}=\frac{\modu{\vec{p}\,}}{2}\naw{1+\vec{x}\,^2}\label{a17}
\end{equation}
\begin{equation}
	\zeta_{i5}=\frac{p_i}{2}\naw{1-\vec{x}\,^2}+\naw{\vec{x}\cdot\vec{p}}x_i\label{a18}
\end{equation}
\begin{equation}
	\zeta_{i6}=\frac{p_i}{2}\naw{1+\vec{x}\,^2}-\naw{\vec{x}\cdot\vec{p}}x_i.\label{a19}
\end{equation}

Some comments concerning eqs.~\eqref{a13}$\div$\eqref{a19} are here in order. It is well known that the Poincare symmetry of massless particles of a given helicity can be extended to the conformal one. Therefore, the set of Darboux variables on relevant coadjoint orbit of Poincare group allows to construct also the additional dilatation and conformal generators. The construction is further simplified by the fact that it is sufficient to deal with the orbit corresponding to vanishing helicity. In this way we obtain the parametrization \eqref{a13}$\div$\eqref{a19}. It is straightforward to check that it obeys the basic Poisson brackets \eqref{a1}. 
Now, according to \cite{Onofri}, \cite{Gyorgyi}, the Hamiltonian of the Kepler problem can be written in terms of $\zeta_{ab}$ as follows:
\begin{equation}
	\mathcal{H}=-\frac{mk^2}{2\zeta_{06}^2}\label{a20}
\end{equation}
or, using eq.~\eqref{a17}
\begin{equation}
	\mathcal{H}=-\frac{2mk^2}{\vec{p}\,^2(1+\vec{x}\,^2)^2}=-\frac{2mk^2}{H}\label{a21}
\end{equation}
with
\begin{equation}
	H=\vec{p}\,^2(1+\vec{x}\,^2)^2.\label{a22}
\end{equation}
Let us make a very simple but useful general remark. Assume that $H$ is some Hamiltonian while $\mathcal{H}=f(H)$ - an arbitrary function of it. Let $(\underline{q}(t),\underline{p}(t))$ be any solution to the Hamiltonian equations of motion for $H$, with $E$ being the corresponding total energy. Then $(\underline{q}(\omega t),\underline{p}(\omega t))$, with $\omega\equiv\frac{d \mathcal{H}}{dH}\rvert_{H=E}$ is a solution to the Hamiltonian equations for $\mathcal{H}$. In other words, both sets of solutions are related by merely rescaling time by a constant along trajectory, energy dependent factor. This, in turn, implies that the trajectories, viewed as the curves in phase space, coincide. Moreover, all integrals of motion which do not depend explicitly on time, coincide as well. 

This is because, in the Hamiltonian formalism, the dynamical variables depend only on the points in phase space (i.e.~on the points on trajectories); as a result an integral of motion is a function constant over any curve in phase space representing trajectory. On the other hand, in Lagrangian formulation the dynamical variables are not only the functions of points on configuration space but also depend on tangent vectors to trajectories.

Most questions concerning Kepler dynamics can be addressed by referring to the Hamiltonian \eqref{a22}. Due to the Poisson commutation rule,
\begin{equation}
	\poisson{\zeta_{\mu\nu},\zeta_{06}}=0,\quad \mu,\nu=1,2,3,5\label{a23}
\end{equation} 
we have six integrals of motion. Three of them,
\begin{equation}
	\zeta_{ij}=x_ip_j-x_jp_i
\end{equation}
or, in standard notation
\begin{equation}
	L_i\equiv\varepsilon_{ijk}x_jp_k \label{a25}
\end{equation}
are the components of angular momentum. The remaining three
\begin{equation}
	A_i\equiv \zeta_{i5}=\frac{p_i}{2}(1-\vec{x}\,^2)+(\vec{x}\cdot\vec{p}\,)x_i=\frac{p_i}{2}(1+\vec{x}\,^2)+(\vec{x}\times\vec{L})_i \label{a26}
\end{equation}
form the components of the counterpart of Runge-Lenz vector. Obviously, $\vec{L}$ and $\vec{A}$ ($\equiv\{\zeta_{\mu\nu}\}$) span, with respect to the Poisson brackets, $SO(4)$ Lie algebra.
Moreover, due to the fact that we are considering the (nongeneric) orbit, they obey the additional relations
\begin{equation}
	\vec{A}\cdot\vec{L}=0\label{a27} 
\end{equation}
\begin{equation}
	\vec{A}\,^2+\vec{L}\,^2=\frac{1}{4}H. \label{a28}
\end{equation}

\section{Nonlinear realizations and chiral dynamics}

Let us note that all conserved quantities $\vec{L}$, $\vec{A}$ are linear in momenta. Therefore, viewed as the generators of canonical symmetry transformations, they actually generate point transformations. In order to analyze them in more detail let us pass to the Lagrangian formalism. The Lagrangian corresponding to the Hamiltonian \eqref{a22} reads
\begin{equation}
	\mathcal{L}=\frac{\dot{\vec{x}}\,^2}{4(1+\vec{x}\,^2)^2}.\label{a29}
\end{equation}
It exhibits, via Noether theorem, the following point symmetries
\begin{itemize}
	\item[-] rotations generated by $G=\delta\vec\varphi\cdot\vec{L}$,
	\begin{equation}
		\delta\vec{x}=\{\vec{x},G\}=\delta\vec{\varphi}\times\vec{x}
	\end{equation}
\item[-] nonlinear transformations generated by $G=\delta\vec{a}\cdot\vec{A}$,
\begin{equation}
	\delta\vec{x}=\{\vec{x},G\}=\frac{1}{2}(1-\vec{x}\,^2)\delta\vec{a}+(\vec{x}\cdot\delta\vec{a})\vec{x}. \label{a31}
\end{equation}
\end{itemize}
It is easy to see that the above nonlinear action of $SO(4)$ on the configuration space fits perfectly into the general scheme of nonlinear realiztion \cite{Coleman}, \cite{Callan}. In fact, locally $SO(4)\sim SU(2)\times SU(2)$ while the rotation group is locally isomorphic to diagonal subgroup $(SU(2)\times SU(2))_{diag}$. The action of $SO(4)$ linearizes on the rotation subgroup and, as we shall see, the components of $\vec{x}$ are preferred (or Goldstone) variables in terminology of Refs.~\cite{Coleman}, \cite{Callan}.

To see this let us note that the elements of $SU(2)\times SU(2)$ may be represented as the pairs $(U,W)$ of $SU(2)$ matrices $U$, $W$ while the diagonal subgroup consists of the pairs $(U,U)$; the relevant coset space may be viewed as the set of pairs $(V,V^+)$. It is sufficient to consider the action of $SU(2)\times SU(2)$ elements which do not belong to the diagonal subgroup. Following Ref.~\cite{Coleman} we write
\begin{equation}
	(U,U^+)\cdot(V,V^+)=(V',V'^+)\cdot(U',U')\label{a32}
\end{equation}
which yields
\begin{equation}
	UV^2U=V'^2.
	\label{a33}
\end{equation}
Let us parametrize the elements $V$ defining the coset manifold as
\begin{equation}
	V=\frac{1}{\sqrt{1+\vec{x}\,^2}}\sigma_0+\frac{i\cdot\vec{x}\cdot\vec{\sigma}}{\sqrt{1+\vec{x}\,^2}} \label{a34}
\end{equation}
with $\sigma_0=\mathbbm{1}$ and $\vec{\sigma}$ being Pauli matrices. Consider the infinitesimal transformations
\begin{equation}
	U=e^{i\delta\vec{a}\cdot\frac{\vec{\sigma}}{2}}\simeq\sigma_0+\frac{i}{2}\delta\vec{a}\cdot\vec{\sigma}. \label{a35}
\end{equation}
By inserting eqs.~\eqref{a34} and \eqref{a35} into \eqref{a33} we find that the transformation rule for $\vec{x}$ coincides with that given by eq.~\eqref{a31}.
 
The Lagrangian \eqref{a29} can be also obtained following the prescription of Refs.~\cite{Coleman}, \cite{Callan}. In fact, the Cartan form restricted to the coset manifold,
\begin{equation}
	\eta = (V^+,V)(dV,dV^+)=(V^+dV,VdV^+)\label{a36}
\end{equation}
takes, in the parametrization \eqref{a34}, the following form
\begin{equation}
	\eta=i\naw{\naw{\frac{2d\vec{x}}{1+\vec{x}\,^2}+\frac{2\vec{x}\times d\vec{x}}{1+\vec{x}\,^2}}\frac{\vec{\sigma}}{2},\naw{-\frac{2d\vec{x}}{1+\vec{x}\,^2}+\frac{2\vec{x}\times d\vec{x}}{1+\vec{x}\,^2}}\frac{\vec{\sigma}}{2}}. \label{a37}
\end{equation}
Taking into account that the generators corresponding to the coset manifold can be chosen as $\naw{\frac{\vec{\sigma}}{2},-\frac{\vec{\sigma}}{2}}$ we conclude that the invariant Lagrangian should be constructed as a function of
\begin{equation}
	\frac{\vec{\eta}}{dt}=\frac{2\dot{\vec{x}}}{1+\vec{x}\,^2} \label{a38}
\end{equation}
invariant under the action of diagonal subgroup, i.e. under rotations \cite{Coleman}, \cite{Callan}. The simplest choice is $\mathcal{L}\sim\naw{\frac{\vec{\eta}}{dt}}^2$ which yields eq.~\eqref{a29}. The momentum components transform linearly (with the coefficients depending on $\vec{x}$). Therefore, according to \cite{Coleman}, \cite{Callan}, they are the so called adjoint variables. In fact, it is not difficult to show that the variables 
\begin{equation}
	\pi_i\equiv \ln(1+\vec{x}\,^2)p_i \label{a39}
\end{equation}
under the action of $(U,U^+)$ undergo the rotation determined by the element $U'\in SU(2)$ entering the right hand side of eq.~\eqref{a32}. Again, this fits nicely into the general scheme of Refs.~\cite{Coleman},\cite{Callan}.

Let us note in passing that the dynamics we are considering is nothing but the dimensional reduction of chiral effecive dynamics of meson isotriplet. Had we replaced in the Lagrangian \eqref{a29} the variable $\vec{x}$ by field variable $\vec{\phi}(x^\mu)$ we would have obtained the effective Lagrangian describing low energy dynamics of pions within so called PCAC scheme \cite{Weinberg}. In fact, in the limit of vanishing light quarks massess the chiral symmetry $SU(2)_L\times SU(2)_R$ emerges which is assumed to be spontaneously broken down to diagonal isovector $SU(2)$ symmetry with pions being the Goldstone degrees of freedom (our parametrization coincides with that used in \cite{Weinberg}, eq. (19.5.18); it is, however, well known that the on-shell amplitudes are, under mild assumption, reparametrization invariant so alternative parametrizations could be used as well).

One can also adopt the geometric point of view. Due to
 $^{SU(2)\times SU(2)}/_{(SU(2)\times SU(2))_{diag}}\sim \,^{SO(4)}/_{SO(3)}\sim S^3$ the Cartan form
\begin{equation}
	\vec{\eta}\sim \frac{d\vec{x}}{1+\vec{x}\,^2} \label{a40}
\end{equation}
defines the $SO(4)$ invariant metric on $S^3$,
\begin{equation}
	ds^2=\vec{\eta}\,^2.\label{a41} 
\end{equation}
This is the starting point of the approach considered in \cite{Keane}.

\section{The canonical transformation}
Up to now we analyzed the properties of dynamics generated by the Hamiltonian $H$, \eqref{a22}. As we argued this provides us complete information about the Hamiltonian $\mathcal{H}$ defined by eq.~\eqref{a21}. On the other hand, the latter is the Kepler Hamiltonian expressed in nonstandard canonical coordinates. We could pass to the standard formulation by Bacry-Gyorgyi transformation \cite{Gyorgyi}, \cite{Onofri}. However, it is advantegous to consider the relevant transformation directly. It is convenient to pass to the spherical coordinates $(r,\theta,\varphi)$. Then the Hamiltonian $H$ reads
\begin{equation}
	H=\naw{p_r^2+\frac{1}{r^2}\naw{p_\theta^2+\frac{p_\varphi^2}{\sin^2\theta}}}\naw{1+r^2}^2. \label{a42}
\end{equation}
The action variables are \cite{Goldstein}:
\begin{equation}
	\mathcal{I}_\varphi\equiv \frac{1}{2\pi}\int\limits_{0}^{2\pi}p_\varphi\,d\varphi=p_\varphi\equiv L_3 \label{a43}
\end{equation}
\begin{equation}
	\mathcal{I}_\theta\equiv\frac{1}{\pi}\int\limits_{\theta_{min}}^{\theta_{max}}\sqrt{\vec{L}\,^2-\frac{p^2_\varphi}{\sin^2\theta}}\,d\theta+p_\varphi=\lvert\vec{L}\rvert \label{a44}
\end{equation}
\begin{equation}
\mathcal{I}_r=\frac{1}{\pi}\int\limits_{r_{min}}^{r_{max}}\sqrt{\frac{E}{(1+r^2)^2}-\frac{\vec{L}\,^2}{r^2}}\, dr=\frac{1}{2}\sqrt{E}-\lvert\vec{L}\rvert=\frac{1}{2}\sqrt{E}-\mathcal{I}_\theta \label{a45}
\end{equation}
which implies
\begin{equation}
	H=4(\mathcal{I}_r+\mathcal{I}_\theta)^2 \label{a46}
\end{equation}
or
\begin{equation}
	\mathcal{H}=-\frac{mk^2}{2(\mathcal{I}_r+\mathcal{I}_\theta)^2}. \label{a47}
\end{equation}
The form of the Kepler Hamiltonian in terms of action variables coincides with the one obtained from the standard approach \cite{Goldstein}; this is nontrivial conclusion because for superintegrable systems action-angle variables are not defined uniquely. Therefore, the canonical transformation relating standard canonical variables to those corresponding to the Hamiltonian $\mathcal{H}$ given by eq.~\eqref{a21} can be found by composing the transformations from both sets of variables to common action-angle ones.

Let us first construct the angle variables for the Hamiltonian \eqref{a22}. The generating function for the relevant transformation reads
\begin{equation}
	S(r,\theta,\varphi;\mathcal{I}_r,\mathcal{I}_\theta,\mathcal{I}_\varphi)=S_r(r;\mathcal{I}_r,\mathcal{I}_\theta)+S_\theta(\theta;\mathcal{I}_\theta,\mathcal{I}_\varphi)+S_\varphi(\varphi;\mathcal{I}_\varphi) \label{a48} 
\end{equation} 
\begin{equation}
	S_r(r;\mathcal{I}_r,\mathcal{I}_\theta)=\int\limits^r\sqrt{\frac{E}{\naw{1+r^2}^2}-\frac{\mathcal{I}_\theta^2}{r^2}}\, dr \label{a49}
\end{equation}
\begin{equation}
	S_\theta(\theta;\mathcal{I}_\theta,\mathcal{I}_\varphi)=\int\limits^\theta\sqrt{{\vec{L}}^2-\frac{\mathcal{I}_\varphi^2}{\sin^2\theta}}\, d\theta \label{a50}
\end{equation}
\begin{equation}
	S_\varphi(\varphi;\mathcal{I}_\varphi)=\int\limits^\varphi p_\varphi\,d\varphi=\mathcal{I}_\varphi\cdot\varphi \label{a51}
\end{equation}
and defines the angle variable through
\begin{equation}
	\alpha_\varphi=\frac{\partial S}{\partial \mathcal{I}_\varphi}=\frac{\partial S_\theta}{\partial\mathcal{I}_\varphi}+\varphi \label{a52}
\end{equation}
\begin{equation}
	\alpha_\theta=\frac{\partial S}{\partial \mathcal{I}_\theta}=\frac{\partial S_r}{\partial\mathcal{I}_\theta}+\frac{\partial S_\theta}{\partial\mathcal{I}_\theta}\label{a53}
\end{equation}
\begin{equation}
	\alpha_r=\frac{\partial S}{\partial\mathcal{I}_r}=\frac{\partial S_r}{\partial\mathcal{I}_r}. \label{a54}
\end{equation}
Eqs.~\eqref{a48}$\div$\eqref{a54} yield
\begin{equation}
	\alpha_\varphi=\varphi+\frac{1}{2}\arcsin\naw{\frac{\mathcal{I}_\varphi^2-\mathcal{I}_\theta^2(1-\cos\theta)}{\mathcal{I}_\theta(1-\cos\theta)\sqrt{\mathcal{I}_\theta^2-\mathcal{I}_\varphi^2}}}+\frac{1}{2}\arcsin\naw{\frac{\mathcal{I}_\varphi^2-\mathcal{I}_\theta^2(1+\cos\theta)}{\mathcal{I}_\theta(1+\cos\theta)\sqrt{\mathcal{I}_\theta^2-\mathcal{I}_\varphi^2}}}
\end{equation}
\begin{equation}
	\begin{split}
	\alpha_\theta=&\alpha_r+\frac{1}{2}\arcsin\naw{\frac{2(\mathcal{I}_r+\mathcal{I}_\theta)^2-\mathcal{I}_\theta^2(1+r^2)}{2(\mathcal{I}_r+\mathcal{I}_\theta)\sqrt{\mathcal{I}_r(\mathcal{I}_r+2\mathcal{I}_\theta)}}}+\\
	&-\frac{1}{2}\arcsin\naw{\frac{2(\mathcal{I}_r+\mathcal{I}_\theta)^2r^2-\mathcal{I}_\theta^2(1+r^2)}{2r^2(\mathcal{I}_r+\mathcal{I}_\theta)\sqrt{\mathcal{I}_r(\mathcal{I}_r+2\mathcal{I}_\theta)}}}-\arcsin\naw{\frac{\mathcal{I}_\theta\cos\theta}{\sqrt{\mathcal{I}_\theta^2-\mathcal{I}_\varphi^2}}} \label{a56}
	\end{split}
\end{equation}
\begin{equation}
	\alpha_r=\arcsin\naw{\frac{(\mathcal{I}_r+\mathcal{I}_\theta)(r^2-1)}{\sqrt{\mathcal{I}_r(\mathcal{I}_r+2\mathcal{I}_\theta)}(r^2+1)}}. \label{a57}
\end{equation}
This is the set of nested equations which can be solved for $r$, $\theta$ and $\varphi$ sequentially, starting from eq.~\eqref{a57}; then $p_\varphi$, $p_\theta$ and $p_r$ can be computed. In this way we obtain the map: $(\alpha_r,\alpha_\theta,\alpha_\varphi,\mathcal{I}_r,\mathcal{I}_\theta,\mathcal{I}_\varphi)\rightarrow(r,\theta,\varphi,p_r,p_\theta,p_\varphi)$. On the other hand, denoting by $(\overline{r},\overline{\theta},\overline{\varphi},\overline{p}_r,\overline{p}_\theta,\overline{p}_\varphi)$ the canonical coordinates within standard approach we have
\begin{equation}
	\mathcal{H}=\frac{1}{2m}\naw{\overline{p}_r^{\,2}+\frac{1}{\overline{r}^{\,2}}\naw{\overline{p}_\theta^{\,2}+\frac{\overline{p}_\varphi^{\,2}}{\sin^2\overline{\theta}}}}-\frac{k}{\overline{r}}. \label{a58}
\end{equation}
The transformation $(\overline{r},\overline{\theta},\overline{\varphi},\overline{p}_r,\overline{p}_\theta,\overline{p}_\varphi)\rightarrow(\alpha_r,\alpha_\theta,\alpha_\varphi,\mathcal{I}_r,\mathcal{I}_\theta,\mathcal{I}_\varphi) $ can be found in many textbooks \cite{Goldstein}. By composing these two maps we find the explicit form of canonical transformation
\begin{equation}
(\overline{r},\overline{\theta},\overline{\varphi},\overline{p}_r,\overline{p}_\theta,\overline{p}_\varphi)\rightarrow(r,\theta,\varphi,p_r,p_\theta,p_\varphi). \label{a59}
\end{equation}
However, let us note that the inverse transformation cannot be obtained explicitly. This is due to the fact that one of the equations relating standard variables to the action-angle ones is transcendental; essentially, it is the Kepler equation which in the standard approach determines time dependence of the radial coordinate.

\section{Conclusions}
The dynamical  group of the (regularized) Kepler problem is $SO(4,2)$. Therefore, the relevant dynamics can be described within the Hamiltonian framework based on the notion of coadjoint orbit of $SO(4,2)$. However, the phase space relevant for Kepler problem is six-dimensional while the generic orbits of $SO(4,2)$ are twelve-dimensional. Consequently, the orbit we have to consider is nongeneric (singular) one. It appears that such an orbit carries the dynamics of relativistic massless point particle with vanishing helicity. The $SO(4,2)$ symmetry of this dynamics is the standard conformal symmetry. It appears that the generators of conformal symmetry may be expressed in terms of properly constructed global (but only for vanishing helicity) Darboux coordinates. In terms of these variables the Kepler dynamics acquires a very simple form. The symmetry transformations generated by the conserved quantities (the angular momentum and Runge-Lenz vector) appear to be the point symmetries. Actually, we arrive at the nonlinear realization of $SO(4)$ linearizing on $SO(3)$ subgroup. The relevant Lagrangian may be viewed as arising from dimensional reduction of the effective Lagrangian describing low energy meson scattering within PCAC scheme in the limit of vanishing masses of light quarks. The price one has to pay for having this nice picture is that the canonical transformation relating the old and new Darboux coordinates is rather complicated. However, only one transcendental equation is involved here which is basically the Kepler equation determining time dependence of radial variable. 

Let us note that the two-dimensional Kepler dynamics has been considered from a similar perspective in Ref.~\cite{Gonera}. There the action-angle variables have been used directly to classify the nonlinear action of symmetry group ($SU(2)$ in this case) according to the general scheme of Coleman et al. \cite{Coleman}, \cite{Callan}.

Finally, it is worth to mention that the relation between free relativistic particle and the Kepler system has been also studied in connection with the idea of "two-time physics" \cite{Bars1}, \cite{Bars2}.

\subsection*{Acknowledgement}
We are grateful to Profs.~Krzysztof Andrzejewski and Pawe\l{} Ma\'slanka for helpful discussion and useful suggestions. This paper was supported by the IDUB grant, Decision No 54/2021.

\end{document}